# Solvatochromic microcavity ion sensors

SHUOYING ZHAO,[1] XIAOLU CUI,[1] JIYANG MA,[1] HAO WU,[1,*] AND QING ZHAO,[1,*]

[1]*Center for Quantum Technology Research, School of Physics, Beijing Institute of Technology, Beijing 100081, China*
**hao.wu@bit.edu.cn*
**qzhaoyuping@bit.edu.cn*

**Abstract:** A novel copper(II) ion ($Cu^{2+}$) sensor has been developed, achieving both a broad dynamic range and high sensitivity, effectively addressing the limitations of conventional fluorescence-based detection methods. The sensor exploits the hydrolysis reaction between rhodamine B hydrazide (RBH) and $Cu^{2+}$, with optical signal amplification enabled by a whispering-gallery-mode (WGM) microcavity laser. By adjusting the solvent type and utilizing the solvatochromic effect, the detection range of the $Cu^{2+}$ and RBH reaction was optimized. The sensor enables rapid, broad dynamic range detection of $Cu^{2+}$, spanning from 5 μM to 3 mM, with a high selectivity, which holds significant potential for future applications in environmental monitoring and biomedical fields.

## 1. Introduction

Copper(II) ions ($Cu^{2+}$) are essential trace elements in living organisms and play key roles in iron absorption, heme synthesis, and the catalysis of various important enzymes [1,2]. However, abnormal fluctuations in $Cu^{2+}$ concentrations can lead to severe health issues. Excessive $Cu^{2+}$ is closely associated with the development of neurodegenerative diseases, such as Wilson's disease, Alzheimer's disease, and Parkinson's disease, and may also cause liver and kidney damage [3–6]. Furthermore, as an environmental pollutant, $Cu^{2+}$ enters water and soil through industrial discharges and agricultural runoff, posing potential threats to both ecosystems and human health [7–10]. Therefore, the efficient detection of $Cu^{2+}$ is essential for clinical diagnostics and environmental monitoring.

Current methods for $Cu^{2+}$ detection include atomic absorption spectroscopy (AAS), inductively coupled plasma mass spectrometry (ICP-MS), and electrochemical techniques, which provide high sensitivity and accuracy. However, the requirements for expensive equipment and complex procedures make rapid on-site detection challenging [11–14]. To overcome these limitations, low-cost fluorescence probes are employed for $Cu^{2+}$ detection due to high sensitivity, rapid response, and visible detection capability [15–18]. Rhodamine B hydrazide (RBH), in particular, exhibits significantly enhanced fluorescence in the presence of $Cu^{2+}$, making it an excellent candidate for $Cu^{2+}$ detection [19,20]. In the absence of $Cu^{2+}$, RBH molecules adopt a closed-loop helical structure, which effectively suppresses fluorescence emission, resulting in minimal visible light absorption. Upon the addition of $Cu^{2+}$, $Cu^{2+}$ ions react with the hydrazide group in RBH, disrupting the closed-loop structure and inducing an opening transition, resulting in the formation of a pink fluorescent open-ring product [21,22]. As the concentration of $Cu^{2+}$ increases, the solution gradually changes from colorless to pink, and the fluorescence intensity is significantly enhanced. The degree of color change and fluorescence enhancement is positively correlated with the $Cu^{2+}$ concentration, allowing for quantitative analysis of the $Cu^{2+}$ levels by monitoring changes in absorbance or fluorescence intensity.

Although RBH is effective for $Cu^{2+}$ detection, the influence of solvent environment on the reaction mechanism and detection performance should not be overlooked. Solvents affect not only chemical reactivity and rates but also the polarity and spectral properties of the system [23,24]. As the solvent polarity increases, a blue shift, known as negative solvatochromism, is typically observed. In contrast, enhanced solvent polarity can lead to a redshift, which is referred to as positive solvatochromism [25–29]. For example, 4,4′-bis-dimethylamino fuchsone is orange in nonpolar toluene, turns red in weakly polar acetone, and becomes purplish-red in strongly polar methanol, demonstrating classic positive solvatochromism [30]. Optimizing the solvent conditions significantly enhances the interaction between RBH and $Cu^{2+}$, thereby improving the fluorescence response and further increasing the sensitivity of $Cu^{2+}$ detection.

To further enhance the sensitivity and accuracy of $Cu^{2+}$ detection, this study integrates a whispering gallery mode (WGM) microcavity laser and solvent effects. The WGM microcavity, with its high-quality factor and small mode volume, effectively amplifies the interaction between light and matter, thereby significantly increasing the interaction time and enhancing both signal strength and resolution [31–34]. Exploiting this advantage, microcavity lasers suppress losses via optical resonance by converting weak fluorescence signals into laser signals. This amplification allows previously difficult-to-detect minor fluorescence changes to be revealed through laser signals, thereby significantly improving the detection signal strength, resolution, and sensitivity [35–39]. The combination of the RBH probe and capillary microcavity laser significantly enhances the sensitivity and precision of $Cu^{2+}$ detection, providing an efficient detection platform for trace $Cu^{2+}$ analysis.

In this study, rhodamine B hydrazide (RBH) was employed as a fluorescent probe to selectively bind $Cu^{2+}$ through a hydrolysis reaction (**Fig. 2**(a)), leading to a pronounced enhancement in fluorescence intensity. The key innovation of this study is the optimization of experimental conditions through the solvatochromic effect, combined with WGM microcavity technology, where fluorescence and laser signals complement each other to enable $Cu^{2+}$ detection from 5 μM to 3 mM. The sensor exhibits excellent ion selectivity, thereby eliminating interference from other ions in practical applications, and consequently provides an effective platform for the rapid, sensitive, and broad dynamic range detection of $Cu^{2+}$ in environmental monitoring and biomedical diagnostics.

## 2. Experiment setup and sample preparation method

The experimental setup used to excite and collect the laser beam generated by the sensor in the capillary microcavity laser is shown in **Fig. 1**. The pump laser source is a nanosecond pulsed laser (Dalian Institute of Chemical Physics, China, 10 Hz, wavelength of 532 nm). After focusing the pump light using a convex lens (L1), the light is further focused onto the capillary by an objective lens to excite the sensor.

To avoid interference from the 532 nm pump light, the emitted laser is first passed through a dichroic mirror (D1), which reflects the green pump light and allows the emitted laser to pass. The emitted laser is then split into two paths by a beam splitter (B1): one beam passes through a long-pass filter (F1) and directly reaches a CMOS camera for real-time monitoring and recording; the other beam passes through another long-pass filter (F2), convex lens (L2) and then transmitted via an optical fiber to an optical spectrum analyzer for further spectral analysis. The multiple filtering steps effectively eliminate interference from the pump source, ensuring the most accurate emission spectrum.

An optical spectrum analyzer (Ocean Optics Maya2000 Pro, 1-nm resolution) was used for preliminary localization and analysis of the emission spectrum. For high-precision spectral analysis, a higher-resolution optical spectrum analyzer (Princeton Instrument, HRS750, ProEM 512) was employed for detailed data acquisition.

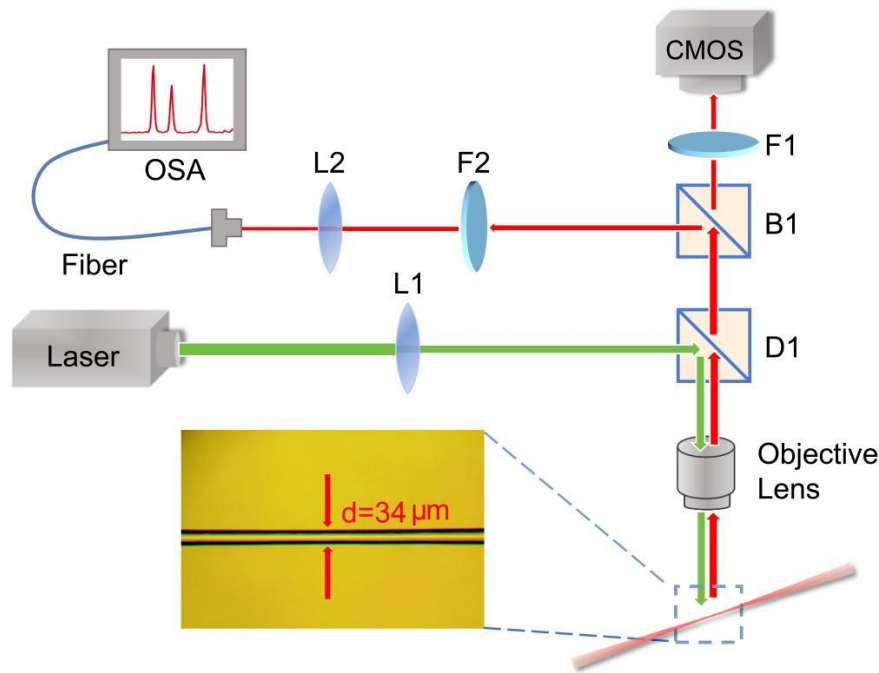

**Fig. 1.** Schematic of experimental setup: L1 and L2 are convex lenses, D1 is a dichroic mirror, B1 is a beam splitter, F1 and F2 are long-pass filters, and OSA is an optical spectrum analyzer. Inset: microscopic image of capillary microcavity.

All chemicals were used as received without further purification. These chemicals include acetone (AR, ≥99.5%; Beijing Tongguang Fine Chemicals Co., Ltd.), acetonitrile (spectroscopic grade, ≥99.8%; MERYER Co., Ltd.), and sodium chloride (NaCl, 99%, Tianjin Heowns Biochemical Technology Co., Ltd.). Methanol (AR, 99.5%) and rhodamine B hydrazide (RBH, 98%) were purchased from Shanghai Aladdin Biochemical Technology Co., Ltd. Anhydrous ethanol (pharmaceutical grade, 99.5%), anhydrous magnesium chloride (99.9%), potassium chloride (99.99%), anhydrous calcium chloride (99.9%), zinc chloride (99.95%), and copper chloride dihydrate (99.99%) were purchased from Shanghai Macklin Biochemical Co., Ltd. Deionized (DI) water was used to prepare all aqueous solutions.

Copper chloride was dissolved in DI water to obtain a 10 mM stock solution. The stock solution was then diluted using stoichiometric calculations to prepare copper chloride solutions with various concentrations within the required range. High-concentration RBH stock solutions were prepared in different solvents, including acetone, anhydrous ethanol, methanol, and acetonitrile, and then diluted to the desired concentrations for experimental use. To eliminate interference from other metal ions, metal perchlorate powders containing $Cu^{2+}$, $Ca^{2+}$, $K^+$, $Zn^{2+}$, $Mg^{2+}$, and $Na^+$ ions were dissolved in DI water to a final concentration of 10 mM. This procedure was used to demonstrate the selectivity of the proposed $Cu^{2+}$ detection method in the experiment.

## 3. Results and discussion

This study systematically characterized the fluorescence and laser spectral properties of a capillary microcavity laser, demonstrating the sensor's feasibility for $Cu^{2+}$ detection. RBH dissolved in acetone was mixed with a fixed concentration of $Cu^{2+}$ and allowed to react completely before being introduced into a capillary microcavity (d = 34 μm) connected to a microfluidic system. The inclusion of the microfluidic system facilitated dynamic circulation, preventing local photobleaching effects from the pump light, thereby enhancing the stability and accuracy of the results. The fluorescence spectrum (black curve) and laser emission spectrum (red curve) were obtained (**Fig. 2**(b)). The fluorescence spectrum exhibited a broad distribution in the 550–650 nm range [full width at half maximum (FWHM) = 15.2 nm], whereas the laser emission spectrum exhibited a sharp peak with a central wavelength of 595 nm (FWHM = 3.4 nm). These two spectra form a distinct contrast.

In addition, by varying the pump light intensity, a series of laser emission spectra as a function of the pump power was obtained, where the direction of the arrow indicates the increase in the pump light energy (**Fig. 2**(c)). At a central wavelength of 592 nm, the FWHM was 0.012 nm. The calculated quality factor exceeds 49000, indicating that the high-Q microcavity effectively suppresses light field losses and supports a stable laser output. The inset of **Fig. 2**(b) shows that the normalized laser intensity exhibits a nonlinear increase with the pump energy density. The laser exhibits a threshold at 2.2 μJ/mm$^2$, providing clear evidence of stimulated emission. This sensor enables the detection of $Cu^{2+}$ concentration based on fluorescence intensity, laser intensity, and threshold characteristics, laying the groundwork for the subsequent optimization of the $Cu^{2+}$ detection dynamic range.

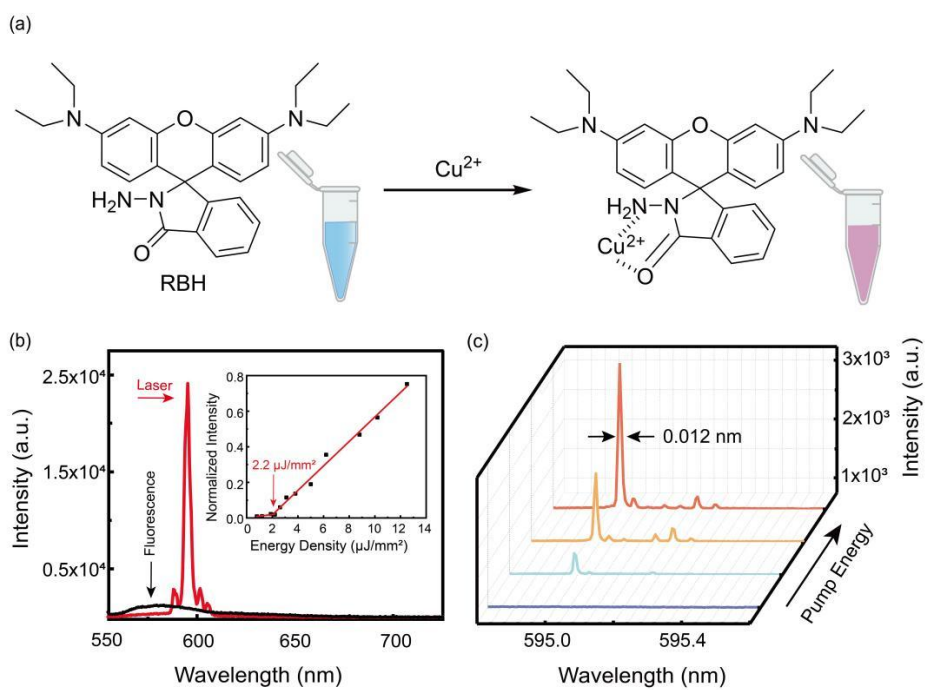

**Fig. 2.** (a) Sensing mechanism of RBH with $Cu^{2+}$. (b) Fluorescence and laser emission spectra. Black curve: fluorescence; red curve: laser emission. Inset: threshold measurement. (c) Laser emission spectra at different pump energies.

Solvent effects are critical in chemical reactions influencing both reaction rate and chemical reactivity. Therefore, selecting an appropriate solvent for RBH as the probe molecule for detecting $Cu^{2+}$ is essential. The absorption spectra of RBH–$Cu^{2+}$ solutions show significant shifts in both wavelength and intensity across different polar solvents (acetone, anhydrous ethanol, methanol, and acetonitrile), which are closely related to the polarity of the solvents (**Fig.**

3(a)). In the 500–580 nm wavelength range, the maximum absorption peak is denoted as $A_0$, whereas, in the 600–800 nm wavelength range, the maximum absorption peak is denoted as $A_1$. Acetone, which has lower polarity, exhibited the largest $A_0$ value, whereas methanol and anhydrous ethanol, which have higher polarity, exhibited the largest $A_1$ values. Acetonitrile and acetone, which have lower polarity, exhibited the smallest $A_1$ values. To further compare the redshift degree of the absorption peaks across the four solvents, the $A_1/A_0$ ratios were calculated for acetone, anhydrous ethanol, methanol, and acetonitrile as solvents; the obtained ratios were 0.011, 0.423, 0.56, and 0.416, respectively. These data indicate that methanol, which exhibits the strongest polarity, exhibits the most significant redshift, whereas acetone, which is the least polar solvent, exhibits absorption peaks in the 500–580 nm range. The positive solvatochromic effect explains this phenomenon, where a redshift occurs with increasing solvent polarity. This is also the reason why the RBH–$Cu^{2+}$ solutions appears purplish-red when methanol and anhydrous ethanol are used as solvents and pale red-pink when acetone and acetonitrile are used (**Fig. 3**(a), inset).

**Fig. 3**(b) shows the laser emission spectra of RBH–$Cu^{2+}$ solutions after being introduced into a capillary microcavity and excited by a 532 nm pump light source. The results show that acetone produces the highest laser intensity, indicating that it effectively promotes the reaction between RBH and $Cu^{2+}$. In contrast, acetonitrile exhibits the lowest laser response. The distribution of the laser intensity across different solvents is consistent with the ultraviolet absorption peak distribution in the 500–580 nm wavelength range. Considering the characteristics of the various solvents, acetone was selected as the ideal solvent for low-concentration $Cu^{2+}$ detection. Although acetonitrile exhibits a weaker laser response, its lower laser intensity makes it suitable for detecting high $Cu^{2+}$ concentrations.

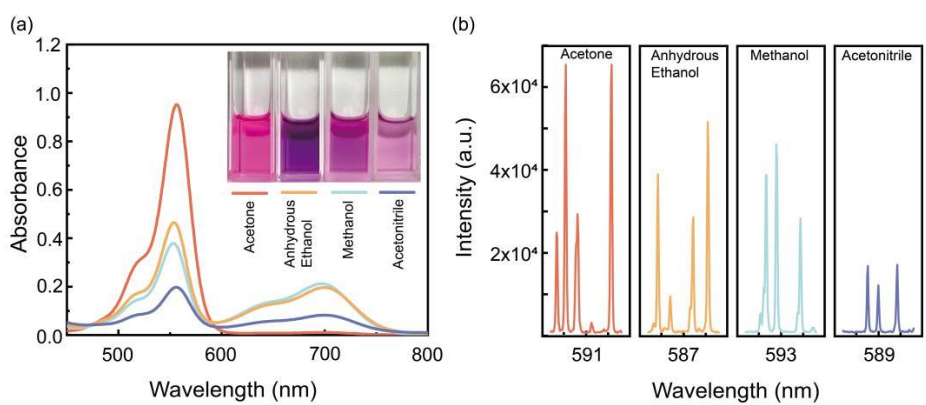

**Fig. 3.** (a) Absorption spectra of RBH–$Cu^{2+}$ solutions in different solvents. Inset: color changes of the solutions. (b) Laser spectra in different solvents under the same pump energy density.

Laser emission spectra of the capillary microcavity sensor were measured in acetone at low $Cu^{2+}$ concentrations (**Fig. 4**(a)). The black arrow denotes the increase in $Cu^{2+}$ concentration from 5 to 35 μM. As the $Cu^{2+}$ concentration increased, the laser emission peak intensity exhibited a pronounced enhancement, demonstrating a positive correlation between $Cu^{2+}$ concentration and emission intensity. The gradual color transition of the solution from light pink to deep pink further corroborates this observation (**Fig. 4**(c), inset). No significant redshift or blueshift of the peak was detected, confirming the spectral stability and ensuring reliable determination of the peak position. The emission peak at 591.95 nm was normalized, and a linear dependence of laser intensity on $Cu^{2+}$ concentration (5–35 μM) is shown in **Fig. 4**(b). Linear fitting yielded a sensitivity of 0.03164 normalized intensity (a.u.) per μM, with a limit of detection (LOD) of 5.04 μM, indicating excellent sensitivity for low-concentration detection.

The threshold of the capillary microcavity laser was measured across varying $Cu^{2+}$ concentrations (**Fig. 4**(c)). As the $Cu^{2+}$ concentration increased, the laser threshold gradually decreased, indicating that higher $Cu^{2+}$ levels required lower excitation energy. This dependence of the laser threshold on $Cu^{2+}$ concentration highlights the potential of the capillary microcavity laser for sensitive $Cu^{2+}$ detection. These results demonstrate that, when acetone is used as the solvent, the sensor can accurately respond to low $Cu^{2+}$ concentrations.

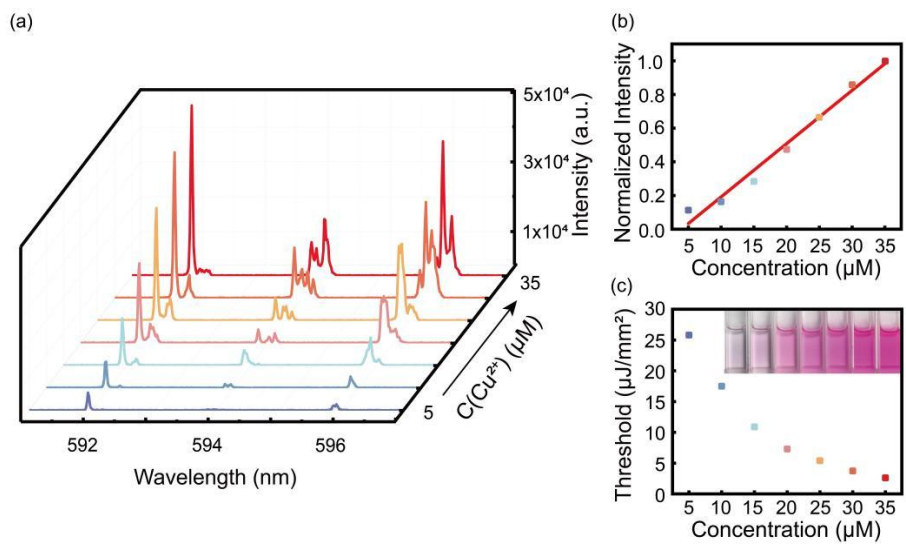

**Fig. 4**. (a) Laser spectra at varying $Cu^{2+}$ concentrations under the same pump energy density. (b) Normalized laser intensity at the 591.95 nm emission peak as a function of $Cu^{2+}$ concentration. (c) Laser threshold variation with $Cu^{2+}$ concentration. Inset: color changes of the solutions. (All experiments used acetone as the solvent.)

To investigate the spectral characteristics of high $Cu^{2+}$ concentrations (50–300 μM), acetonitrile was used as the solvent. Under a fixed pump energy, the emission spectra at different concentrations revealed a clear enhancement of laser intensity with increasing $Cu^{2+}$ concentration (**Fig. 5**(a)). Quantitative analysis at the 587.4 nm emission peak (**Fig. 5**(b)) showed a strong linear correlation between laser intensity and $Cu^{2+}$ concentration, with a fitted coefficient of determination ($R^2$) of 0.995, confirming the sensor's robust linear response within this range. The calculated LOD of the sensor was approximately 27 μM, further demonstrating the sensor's effectiveness for high-concentration detection. Moreover, as the $Cu^{2+}$ concentration increased from 50 to 300 μM, the lasing threshold decreased markedly from 19.3 to 4.1 μJ/mm² (**Fig. 5**c), indicating that higher $Cu^{2+}$ concentrations reduce the excitation energy required for lasing. Overall, these results highlight that, with acetonitrile as the solvent, the capillary microcavity laser provides reliable and quantitative detection of $Cu^{2+}$ at high concentrations.

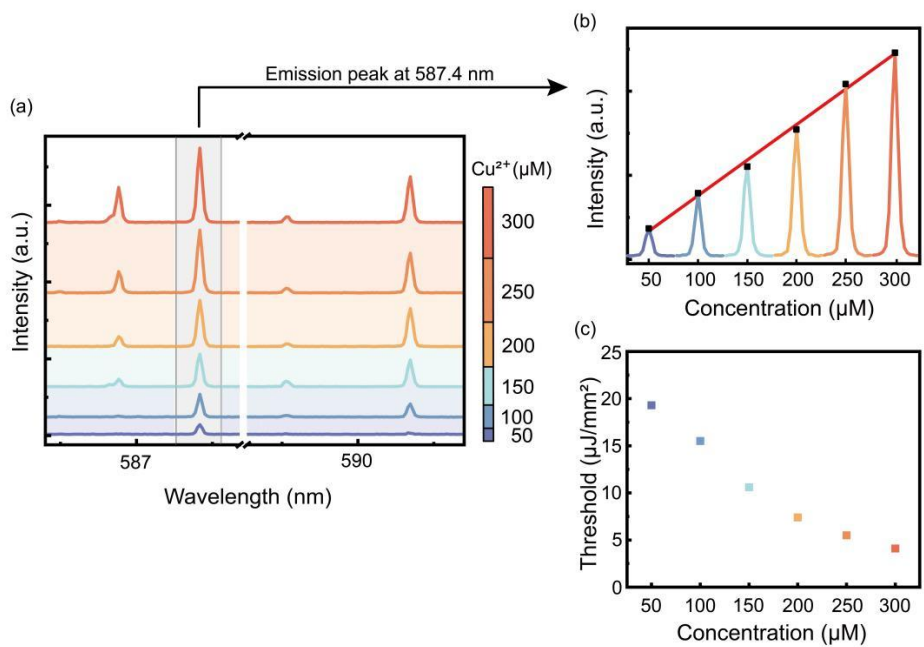

**Fig. 5.** (a) Laser spectra at varying $Cu^{2+}$ concentrations under the same pump energy density. (b) Intensity at the 587.4 nm emission peak as a function of $Cu^{2+}$ concentration. (c) Laser threshold variation with $Cu^{2+}$ concentration. (All experiments used acetonitrile as the solvent.)

Beyond laser detection, fluorescence measurements in acetonitrile are also highly effective for detecting high $Cu^{2+}$ concentrations (0.075–3 mM), effectively addressing the gap in high-concentration detection. The fluorescence response of $Cu^{2+}$ in acetonitrile is shown in **Fig. 6**(a). As the $Cu^{2+}$ concentration increased, the fluorescence emission peak intensity rose significantly, exhibiting a clear concentration dependence accompanied by a redshift. **Fig. 6**(b) presents quantitative analysis of the central wavelength shift and fluorescence intensity. Although a data jump occurs at 0.075 mM, the measurements from 0.2 to 3 mM exhibit a good linear relationship. The left vertical axis shows that the central wavelength shifts from 579 nm at 0.075 mM to 589.5 nm at 3 mM, corresponding to a significant redshift of 10.5 nm. The right vertical axis illustrates the fluorescence intensity as a function of $Cu^{2+}$ concentration, revealing a strong positive correlation. Further analysis indicates that the fluorescence-based LOD of the sensor is approximately 0.145 mM, demonstrating its effective performance for high-concentration $Cu^{2+}$ detection.

Thus, with acetonitrile as the solvent, the sensor exhibited excellent performance in both laser and fluorescence detection of high-concentration $Cu^{2+}$.

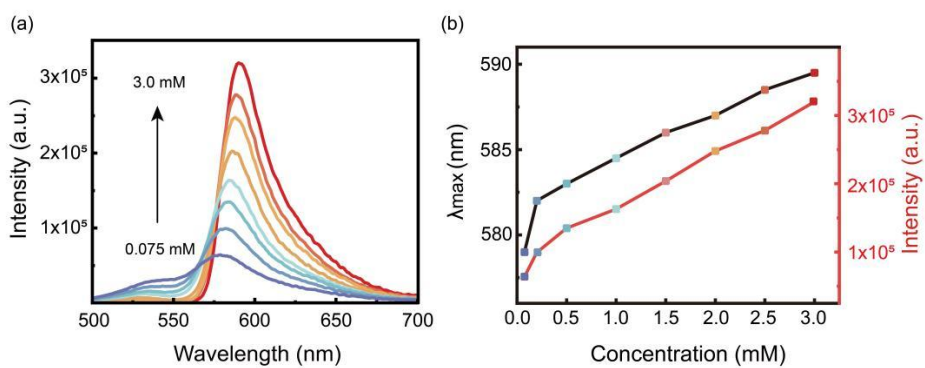

**Fig. 6.** (a) Fluorescence spectra at varying $Cu^{2+}$ concentrations. (b) Central wavelength and fluorescence intensity at varying $Cu^{2+}$ concentrations. (All experiments used acetonitrile as the solvent.)

In practical detection, the sensor's selective response to $Cu^{2+}$ is crucial. The effects of different metal ions ($Cu^{2+}$, $Ca^{2+}$, $K^+$, $Zn^{2+}$, $Mg^{2+}$, and $Na^+$) on the absorbance and laser intensity of the RBH solution were investigated in acetone and acetonitrile (**Fig. 7**). The inset shows the color changes of the solutions after addition of different ions: only the $Cu^{2+}$ solution appeared pink, whereas the solutions containing other metal ions remained colorless. Further analysis revealed that $Cu^{2+}$ exhibited the highest absorbance, while the absorbance of the other metal ions was negligible. In laser detection, only $Cu^{2+}$ induced a laser response; the other metal ions produced no emission, even under high pump energy, indicating the sensor's specific detection capability toward $Cu^{2+}$. These results demonstrate that the sensor possesses high selectivity for $Cu^{2+}$, effectively excluding interference from other metal ions.

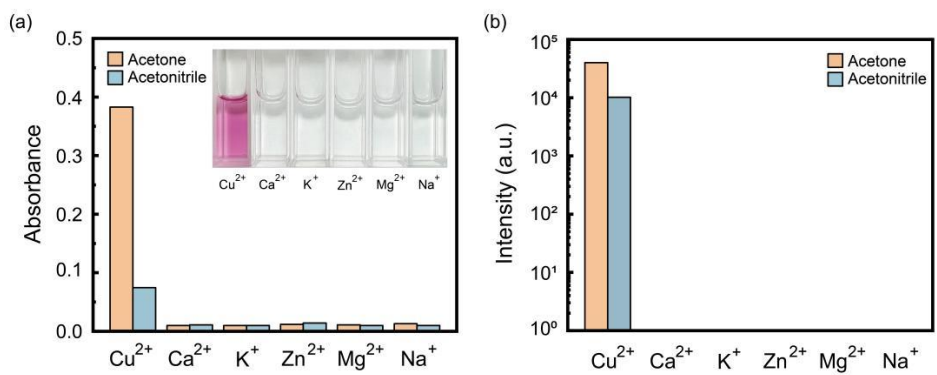

**Fig. 7.** (a) Absorption spectra of solutions containing different metal ions. Inset: color changes of the solutions. (b) Laser spectra of solutions containing different metal ions under the same pump energy density.

## 4. Conclusion

In summary, a novel $Cu^{2+}$ sensor was developed based on the hydrolysis reaction of Rhodamine B hydrazide with $Cu^{2+}$, combined with solvent effects and whispering-gallery-mode (WGM) microcavity-enhanced detection. By leveraging the complementary advantages of acetone and acetonitrile, the solvent effects optimize experimental conditions and broaden the detection range. The co-detection of laser and fluorescence signals enables a broad dynamic range from 5 μM to 3 mM, with high selectivity, specific response to $Cu^{2+}$, low detection limit, and rapid response. This work provides a new approach for $Cu^{2+}$ sensing and holds promising potential for applications in environmental monitoring and related fields.

**Funding.** National Postdoctoral Program for Innovative Talents (BX20200057).

**Disclosures.** The authors declare no conflicts of interest.

**Data availability.** Data underlying the results presented in this paper are not publicly available at this time but may be obtained from the authors upon reasonable request.